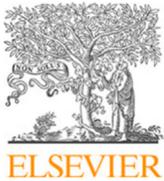
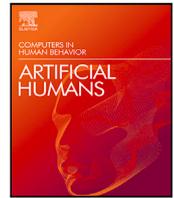

Research pape

# How voice and helpfulness shape perceptions in human–agent teams

Samuel Westby [a,*], Richard J. Radke [b], Christoph Riedl [a], Brooke Foucault Welles [a]

[a] *Northeastern University, 117 Huntington Ave #1010, Boston, MA 02115, USA*
[b] *Rensselaer Polytechnic Institute, 110 8th Street, Troy, NY, 12180, USA*

ABSTRACT

Voice assistants are increasingly prevalent, from personal devices to team environments. This study explores how voice type and contribution quality influence human–agent team performance and perceptions of anthropomorphism, animacy, intelligence, and trustworthiness. By manipulating both, we reveal mechanisms of perception and clarify ambiguity in previous work. Our results show that the human resemblance of a voice assistant's voice negatively interacts with the helpfulness of an agent's contribution to flip its effect on perceived anthropomorphism and perceived animacy. This means human teammates interpret the agent's contributions differently depending on its voice. Our study found no significant effect of voice on perceived intelligence, trustworthiness, or team performance. We find differences in these measures are caused by manipulating the helpfulness of an agent. These findings suggest that function matters more than form when designing agents for high-performing human–agent teams, but controlling perceptions of anthropomorphism and animacy can be unpredictable even with high human resemblance.

## 1. Introduction

The evolving relationships between humans and technology have allowed us to augment human capabilities for decades (Bush et al., 1945; Licklider, 1960). Each advancement in technology invites new opportunities. Now, with the increasing capabilities of generative artificial intelligence (AI), we are beginning to scratch the surface of how to apply conversational voice assistants. They help us select songs, maintain our shopping lists, and search for information in real time. These assistants have also made their way into complicated settings such as mental health treatment (Ahmad, Siemon, Gnewuch, & Robra-Bissantz, 2022; Bérubé, Schachner, Keller, Fleisch, v Wangenheim, Barata, & Kowatsch, 2021) and improving a team's collective intelligence (Westby & Riedl, 2023). The combination of a human working cooperatively with a voice assistant on a shared goal forms a human–agent team (HAT) (McNeese, Demir, Cooke, & Myers, 2018).

To reach the full potential of voice assistants and autonomous agents more generally, O'Neill, Flathmann, McNeese, and Salas (2023) call for designers, developers, and researchers to study the mechanisms that link independent variables to dependent variables in human–agent teams. Outside variables may moderate or mediate the effects of specific agent attributes. As agents gain the ability to act in increasingly complex scenarios, understanding higher-order interactions will be necessary to create an effective HAT.

The perfect agent for each HAT is unique. If the agent offers correct solutions to a problem, a human may still prefer their answer (Kawaguchi, 2021). Perceptions such as anthropomorphism and trust can change an agent's acceptance and usage (Hu, Lu, et al., 2021; Wagner, Nimmermann, & Schramm-Klein, 2019). The Computers Are Social Actors (CASA) paradigm (Nass & Moon, 2000) provides a framework to understand and manage human–agent interaction. It states that humans create expectations for all social interactions, including interactions with technology. This simplifies the complexities of social life but can lead to undesirable interactions when social scripts are violated.

When an agent's voice deviates from a listener's expectation, the listener can feel dehumanized perceptions (Harris & Fiske, 2015; McAleer, Todorov, & Belin, 2014). Speakers with foreign accents do not follow the same social scripts as native speakers and are trusted less than speakers with native accents (Lev-Ari & Keysar, 2010). Robotic-sounding voices can be trusted the same as human-sounding voices (Abdulrahman & Richards, 2022), but (Kulms & Kopp, 2019) found increased self-reported trust for human-like agents versus agents with no human resemblance. There is not one clear choice when designing an agent's human resemblance.

An additional factor in agent design is the quality of an agent's contributions. Contribution quality has a range of effects on trust and perception. Erroneous, flawed, and unhelpful contributions can increase perceived anthropomorphism, (Salem, Eyssel, Rohlfing, Kopp, & Joublin, 2013), decrease it (Salem, Lakatos, Amirabdollahian, & Dautenhahn, 2015), or have no effect (Mirnig et al., 2017). This impacts a member's willingness to interact with the agent and the ability of an agent to communicate with the human (Jung, Martelaro, & Hinds, 2015). Anthropomorphism also affects the acceptance of voice agents (Wagner et al., 2019).

In this work, we study the second-order effects between the human resemblance of a voice assistant's voice and the assistant's contributions to a human–agent team. We found that the helpfulness of our agent negatively moderates the effect of voice (human or robotic) on animacy and anthropomorphism. A human-sounding agent making unhelpful contributions is rated as more human than when it makes helpful contributions. The converse is true for a robotic-sounding agent. We also found that the quality of an agent's contribution significantly affected team performance at the three-quarter mark, but was insignificant at the end of the task. Voice type was not significant for both time points. Perhaps learning that an agent teammate was unhelpful allowed teams to progress through the puzzle while ignoring its bad advice. Voice type did not affect perceived intelligence and perceived trustworthiness. Both were entirely affected by the agent's quality of contribution.

Designers and researchers should recognize that perceived anthropomorphism and animacy are not solely determined by an agent's voice resemblance to humans. Its actions significantly influences what is considered "human-like". To increase perceived intelligence and trustworthiness, designers should prioritize improving agent functionality over fine-tuning voice characteristics. While voice may affect initial adoption, users ultimately judge agents based on their ability. For performance-focused teams, optimizing contribution quality should also take precedence.

## 2. Theory development

### 2.1. Human–agent teams

A human–agent team is formed when at least one human works cooperatively with at least one autonomous agent (McNeese et al., 2018). The agent can come in many forms such as a voice assistant, robot, digital avatar, or text (National Academies of Sciences, Engineering, and Medicine et al., 2022). Like human–human teams, the humans and agents assume different roles so the whole can be greater than the sum of its parts. The agent can be an assistant, supervisor, or collaborator (Endsley, 2017). Agents can moderate team conflict (Jung et al., 2015), improve situational awareness (Chen et al., 2013), or process complicated information (Kawaguchi, 2021).

### 2.2. Perceptual constructs

Interactions with agents depend on the user's perceptions of the agent. How trustworthy is the agent? Is the agent useful? Does the agent act like a human? The Computers Are Social Actors (CASA) paradigm explains that human perceptions of agent teammates come from social scripts and behavioral expectations (Gambino, Fox, & Ratan, 2020; Nass & Moon, 2000). We have expectations for each interaction ranging from talking to a friend, logging into a website, or interacting with Siri. Semi-intelligent actors like Siri are less than human and more than objects, which creates ambiguity about how we should interact with them. Which heuristics apply? Which traits matter?

Agents do not fit in the ontological categories of object or human. Instead, we have formed a third category for semi-human actors (Jr et al., 2011). This category is a moving target (Festerling & Siraj, 2022). What is "human" appears to be shrinking, and what is semi-human appears to be growing. For example, creating music is no longer a purely human endeavor. AI music generation models like MusicGen (Copet et al., 2024) can create songs that are nearly indistinguishable from songs created by experienced human musicians.

Perceptions of agents in this semi-human category will become increasingly important as they become more common. Here we focus on perceived anthropomorphism, perceived animacy, perceived intelligence, and perceived trustworthiness.

*Perceived anthropomorphism and animacy.* Anthropomorphization is attributing human traits to non-human entities (Epley, Waytz, & Cacioppo, 2007). To measure this, questionnaires can ask for comparisons such as *fake vs. natural* and *unconscious vs. conscious* (Bartneck, Kulić, Croft, & Zoghbi, 2009). Animacy is the perception of acting on one's own accord (Bartneck et al., 2009). Is the agent *mechanical vs. organic*? Is the agent *apathetic vs. responsive*? This is similar to the concept of agency or autonomy, which is one's right to act on one's will (Shapiro, 2005). Perceived anthropomorphism and perceived animacy are closely related as they measure similar perceptions of "humanness". They both drive bias-based behaviors such as trust, persuasion, and usage (Atkinson, Mayer, & Merrill, 2005; Fogg, 2002). Animacy is less studied, which makes it an interesting measure to include in this study.

*Intelligence.* Great teams predict the thoughts and behaviors of others and integrate those predictions into their actions (DeChurch & Mesmer-Magnus, 2010; Klien, Woods, Bradshaw, Hoffman, & Feltovich, 2004; McNeese et al., 2018). In human–agent teams, the agent must do this to be seen as intelligent and be seen as a teammate (McNeese et al., 2018). To measure perceived intelligence, questionnaires can ask for comparisons such as *incompetent vs. competent* and *ignorant vs. knowledgeable* (Bartneck et al., 2009). Zhang, McNeese, Freeman, and Musick (2021) found that AI experts rated skill/intelligence as the most important attribute in an AI teammate. Perceived intelligence has little relation to perceived anthropomorphism and perceived animacy (Moussawi & Benbunan-Fich, 2021). It generally captures performance-based perceptions (Koda & Maes, 1996).

*Trust.* Trust in agents alters a user's willingness to take an agent's suggestions (Cohen, Demir, Chiou, & Cooke, 2021). If an agent makes mistakes, a designer would want humans to calibrate their trust in the agent. Over-trusting and under-trusting a teammate's contributions have detrimental effects on team performance (Parasuraman & Riley, 1997; Yang, Huang, Scholtz, & Arendt, 2020). Low trust in teammates also limits the collaborative potential. Low-trust teams are less likely to have shared mental models or have a separation of roles (McAllister, 1995). Controlling trust in voice assistants is needed to unlock the full potential of voice agents in the real world (Hu et al., 2021). Trust has been related to measures of humanness (anthropomorphism and animacy) (Chiou, Schroeder, & Craig, 2020; Seaborn, Miyake, Pennefather, & Otake-Matsuura, 2021). In some work, trust is similar to perceived intelligence. It can be entirely determined by agent performance (Dzindolet, Peterson, Pomranky, Pierce, & Beck, 2003) (see Fig. 1).

### 2.3. Perceptual effects

#### 2.3.1. Perceived anthropomorphism

*The effect of human resemblance.* Agents with human voices can be anthropomorphized more than agents with robotic voices (Ferstl, Thomas, Guiard, Ennis, & McDonnell, 2021; Moussawi & Benbunan-Fich, 2021). Voices with more human-like prosodic variability are also perceived as more human (Schroeder & Epley, 2016; Seaborn et al., 2021). Elements such as pace and intonation can indicate human thought and emotion. Other external factors such as a user's loneliness (Wang, 2017) and in-group bias (Eyssel & Kuchenbrandt, 2012) support changes in perceived anthropomorphism.

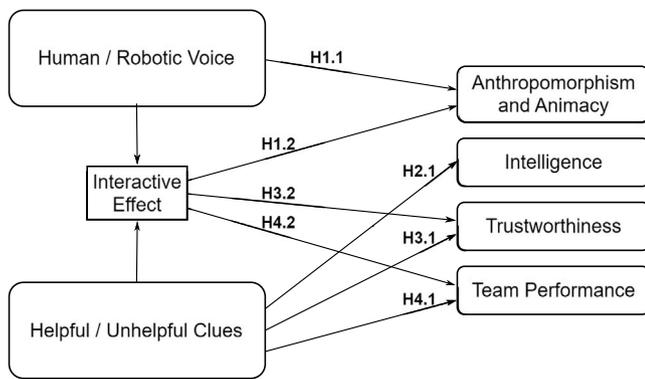

**Fig. 1.** The constructs of this paper and their hypothesized relationships. All hypotheses seek to answer how the voice of a voice assistant combines with the assistant's helpfulness to alter anthropomorphism, animacy, perceived intelligence, trustworthiness, and team performance.

*The effect of contribution quality.* Agent behavior affects perceived anthropomorphism. For example, socially adapted behavior increased perceived anthropomorphism (Wagner & Schramm-Klein, 2019). Errors or abnormal behavior can increase, decrease, or not affect perceived anthropomorphism (Mirnig et al., 2017; Salem et al., 2013, 2015). Developing a relationship with the voice agent is correlated with higher perceived anthropomorphism (Seymour & Van Kleek, 2021).

In Salem et al. (2013), researchers used a Wizard of Oz technique to give tasks to participants through a robot teammate. This robot had three versions: neutral, flawless, and flawed. Neutral had no gestures, flawless had congruent co-verbal gestures, and flawed had incongruent gestures. The instructions and voice remained the same across all three conditions, but the manipulation of gestures affected subjective and objective outcomes. Participants in the flawed condition performed worse than the other two conditions but rated their robot partner highest in anthropomorphism.

In a similarly designed study, researchers again used Wizard of Oz techniques and a robot to ask participants a set of questions and requests (Salem et al., 2015). These questions started as benign as, "Would you like to listen to some music?" with the options, "Yes, Rock", "Yes, Classical", and "No, thanks". In the flawless condition, the robot would obey the participant's choice. In the flawed condition, the robot would take an action different from the participant's choice. Next, the robot asked participants a series of unusual requests such as, "Pour orange juice into the plant on the windowsill". Contrary to Salem et al. (2013), robots with faulty behavior were rated lowest in perceived anthropomorphism, while treatment did not affect participants' behavioral trust in the agent.

To further complicate the effect of an agent's contribution on perceived anthropomorphism, another study found no evidence that contribution quality alters the perceived anthropomorphism of agent teammates (Mirnig et al., 2017). The incongruent results of these three studies suggest that additional variables moderate the effect of an agent's contribution quality on perceived anthropomorphism. Based on works results we have the following hypotheses:

**H1.1** Agents with human voices will be perceived as more anthropomorphic and animated than agents with robotic voices.
**H1.2** Voice type (human vs. robotic) will moderate the effect of contribution quality (helpful vs. unhelpful) on perceived anthropomorphism and perceived animacy.

Fig. 1 shows the relationships between all constructs and hypotheses.

*2.3.2. Perceived intelligence*
In human–human interactions, voice manipulation can bias ratings of intelligence (Hughes, Mogilski, & Harrison, 2014; Schroeder & Epley, 2015). This bias is less understood in human–agent interactions. Moussawi and Benbunan-Fich (2021) found that voice type alone is not correlated with perceived intelligence. Tsiourti, Weiss, Wac, and Vincze (2019) found that mismatched expressions between the body and voice in agents lowered perceived intelligence. Koda and Maes (1996) found that only agent competency, not human resemblance impacted perceived intelligence. A voice assistant with a human voice may subtly affect perceptions of intelligence, but this effect may be diminished in the presence of other factors.

**H2.1** Voice type will have no detectable impact on perceptions of intelligence. Intelligence will be impacted only by the agent's contribution.

*2.3.3. Trust*
*The effect of human resemblance.* Humans are more likely to accept suggestions from an agent with a human voice than from an agent with a robotic voice (Chérif & Lemoine, 2019; Chiou et al., 2020; Schreuter, van der Putten, & Lamers, 2021; Seaborn et al., 2021). More human resemblance in voice is also correlated with higher cognitive- and emotion-based trust (Moussawi & Benbunan-Fich, 2021). While cognitive-based trust may not be associated with a human's intention to use a voice assistant, emotion-based trust has a significant correlation (Moussawi & Benbunan-Fich, 2021). Aside from voice, human-like behaviors such as eye contact or showing group-based emotion make an agent seem more trustworthy (Correia, Mascarenhas, Prada, Melo, & Paiva, 2018; Jung et al., 2015).

*The effect of contribution quality.* Helpful agents that offer quality contributions, regardless of voice or form, should be perceived as trustworthy. Users can assess an agent's true helpfulness (Yin, Wortman Vaughan, & Wallach, 2019). Low-quality interactions with an agent can decrease an individual's trust in that system (Nasirian, Ahmadian, & Lee, 2017). Violations of social rules can decrease the agent's trustworthiness (Pitardi & Marriott, 2021).

*Human resemblance combined with contribution quality.* When combined with contribution quality, the effect of human resemblance may dwindle. In a meta-analysis of factors affecting trust in human–robot interaction, Hancock et al. (2011) found that robot attributes (for example, personality and anthropomorphism) had almost no association with trust when compared with robot performance. Although human voices may be more trustworthy than robotic voices, perceptions of trust may be formed using more objective evidence when evidence is available. From work in psychology, we know perceptions of ability and contribution are subject to many human biases (Argyle & McHenry, 1971; Talamas, Mavor, & Perrett, 2016), but users can overwrite those biases by developing trust according to the agent's utility. More work is needed to parse out the effects of an agent's contribution to the effects of an agent's human resemblance.

Users must calibrate their trust to find a stable point that aligns with the agent's ability (Demir et al., 2021; Lee & See, 2004). Work on trust in automation shows that humans form their initial trust based on subjective biases from context or physical characteristics (Lee & See, 2004). After interacting with the automation, trust becomes more objective and based on ability (Dzindolet et al., 2003; Lee & See, 2004).

(Yin et al., 2019) asked people to make guesses assisted by a machine learning model. They manipulated the stated accuracy and the observed accuracy of the model and found that both manipulations influenced a person's willingness to accept a suggestion. A high stated accuracy but a low observed accuracy will still make humans distrustful of the model. Participants saw past the errors in stated accuracy to uncover the real model accuracy. The participants initially trusted the AI, but the final trust was calibrated to objective model accuracy.

(Kulms & Kopp, 2019) conducted a 3 × 2 between-subjects experiment by varying the agent (computer, virtual, human) and advice quality (good, mixed). They measured HAT performance, behavioral trust, and self-reported trust. Participants played 3 rounds of a web-based puzzle game. After Round 3, there was a significant positive effect of advice quality on performance. Participants also requested more advice from the "good advice" agents, regardless of human resemblance. Participants in the virtual and human agent conditions were not significantly different in behavioral and self-reported trust. Both were significantly higher than the computer condition. This means that human resemblance had some effect on performance and trust, but advice quality outweighed voice type.

(Visser et al., 2016) found that if an agent makes a mistake, a human's trust in a human-like agent may be less affected than a human's trust in an agent with less human resemblance. This means an unhelpful human-like agent may be more detrimental to a HAT than an unhelpful non-human-like agent. According to CASA, abiding by a human's social script would allow an agent to slip past our skeptical safeguards. It can force humans to act on autopilot as if nothing is wrong. Over time, humans are good at learning which agents to trust (Demir et al., 2021; Lee & See, 2004), but a human resemblance could slow that process.

**H3.1** Participants will trust helpful agents more than unhelpful agents.

**H3.2** Participants will trust unhelpful human agents more than unhelpful robotic agents.

*2.4. Performance effects*

*Human resemblance affects performance.* The voice of a voice assistant can impact performance. Atkinson et al. (2005) found that students learned better from an agent with a human voice than from an agent with a machine-synthesized voice. They explained this finding using social agency theory. Similar to CASA, this theory states that individual actions are constrained by social cues, including voice type (Moreno, Mayer, Spires, & Lester, 2001). Students were more willing to treat human-voiced agents like teachers because the human voice satisfied human-like social cues. On the contrary, Craig and Schroeder (2017) more recently found that modern synthesized voice agents were more effective as teachers than classic-voiced or human-voiced agents. Chiou et al. (2020) found no differences in learning outcomes when they varied the voice type of a virtual assistant. The effect of voice type on performance is a moving target (Festerling & Siraj, 2022). As our daily interactions with semi-human actors change, so do the effects of agent attributes on HAT outcomes. Research on the mechanisms of human–agent interaction can help designers confidently choose agent attributes and help the field understand ongoing changes.

*Contribution quality affects performance.* With technology deployed in real-world teams, an autonomous agent will inevitably execute faulty or unhelpful actions. How do these errors affect the performance of a human–agent team? If humans over-trust (or under-trust) their agents, HAT performance can suffer (De Jong & Dirks, 2012; Dzindolet et al., 2003; Lee & See, 2004) and lead to humans not using the agents (Parasuraman & Riley, 1997). If the errors add cognitive strain on the teammates, that may outweigh any benefits the agent provides (Flathmann et al., 2023; Gambino et al., 2020; Paleja, Ghuy, Ranawaka Arachchige, Jensen, & Gombolay, 2021). Oddly, there are situations where participants may like imperfect agents more and are more receptive to their suggestions (Kawaguchi, 2021).

*Human resemblance combined with contribution quality affects performance.* Because a human may trust an agent with a human voice more than an agent with a robotic voice (Visser et al., 2016), teams with an unhelpful human-voiced agent may have a harder time progressing through an activity than teams with an unhelpful robotic-voiced agent. If the activity is long enough, participants should have time to calibrate trust (Demir et al., 2021; Lee & See, 2004). The time it takes to calibrate trust to an unhelpful agent may show up as performance differences between groups. More work is needed to understand how an agent's contribution moderates the effects of an agent's form.

**H4.1** Teams with a helpful agent will perform better than teams with an unhelpful agent, regardless of the agent's voice.

**H4.2** Teams with an unhelpful human-voiced agent will perform worse than teams with an unhelpful robotic-voiced agent.

## 3. Methodology

We conducted a 2 × 2 between-subjects experiment where we manipulated an audio-only agent's vocal humanness (human vs. robotic) and contribution quality (helpful vs. unhelpful). Participants were divided into teams of three or four, collaborating via video conferencing software to solve a puzzle. The audio-only agent offered suggestions and clues for the duration of the task. After the puzzle, participants completed a post-experiment questionnaire and were paid $20/hour in Amazon gift cards for up to 90 min of their time. Fig. 3 shows the experimental flowchart.

*3.1. Participants*

We recruited 69 people from the Northeast United States to participate in this study. They were recruited through flyers, email lists, and word of mouth mostly in a university setting. Eligibility criteria required participants to be fluent in English, over the age of 18, and capable of using a computer. We divided participants into 20 teams of three or four people. That made five teams for each treatment. Team size varied between three and four people based on convenience. Four participants were scheduled for each session, and if three or more attended then the session continued. For one session only one participant attended, and they were given the option to receive payment or be placed in a future team. Participant ages ranged from 18 to 34 (mean 21.7 ± 3.67, Table 1).

While recruitment aimed for diversity in terms of background and ethnicity, the sample was a convenience sample. Participants were predominantly young adults with at least some post-secondary education. Of the participants, 37.7% self-identified as Asian (except South Asian), 27.5% as White, 15.9% as South Asian, 8.7% as Black, 5.8% as Latino-Hispanic, and 4.3% as Other.

While the methods used in this study could indeed be applied to dyads (one human and one agent), we chose to study teams with three to four human participants. This better reflects the complexity and dynamics of real-world collaborative environments. In many workplace settings, tasks often involve groups of three or more people working together (Riedl, Kim, Gupta, Malone, & Woolley, 2021). Additional layers of communication, coordination, and decision-making arise. By studying teams of this size, we aim to capture a broader range of social dynamics that are more representative of multi-person collaborative tasks. We acknowledge that the dynamics observed in three or four-person groups may differ from those in dyads. Future research could explore how the effects vary across different group sizes.

*3.2. The task*

Jay Lorch and Michelle Teague designed the Cursed Treasure Puzzle for the 2005 Microsoft Intern Puzzle Day (https://jaylorch.net/puzzles/CursedTreasure/). The puzzle involves a set of cursed treasure chests; players need to uncover the meanings of each curse to determine a final secret word. For example, the *octopus curse* is present only if no gems in a chest are touching, and absent if any gems are touching (Fig. 2). There are five curses to solve then two additional steps to decode the final word. In total, teams need to pass seven milestones to solve the puzzle in the allotted 40 min. This gives us a granular measure of team performance embedded into a complex and interesting task.

**Table 1**
Participant demographics in each of the four agent treatments: human voice/helpful clues, human voice/unhelpful clues, robotic voice/helpful clues, and robotic voice/unhelpful clues. English fluency is self-reported as Native Proficiency (NP), Full Proficiency (FP), and Proficiency (P). No participants reported a lower fluency.

| Treatment | Participants | Female | Teams | Mean age (SD) | NP | FP | P |
| --- | --- | --- | --- | --- | --- | --- | --- |
| Human/Helpful | 17 | 65% | 5 | 22.2 (4.56) | 65% | 35% | 0% |
| Human/Unhelpful | 17 | 47% | 5 | 21.9 (3.37) | 47% | 53% | 0% |
| Robotic/Helpful | 18 | 56% | 5 | 20.9 (2.78) | 67% | 17% | 17% |
| Robotic/Unhelpful | 17 | 71% | 5 | 21.8 (3.62) | 59% | 41% | 0% |
| All | 69 | 59% | 20 | 21.7 (3.67) | 59% | 36% | 4% |

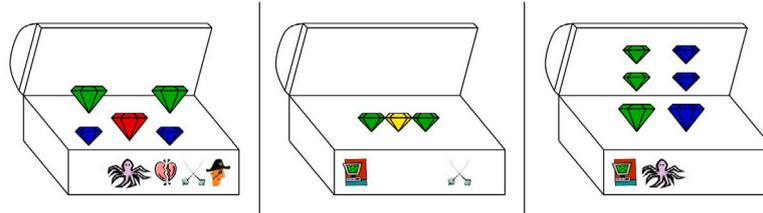

**Fig. 2.** A sample of three chests in the Cursed Treasure puzzle. Notice that the octopus curse relates to the separation of gems.

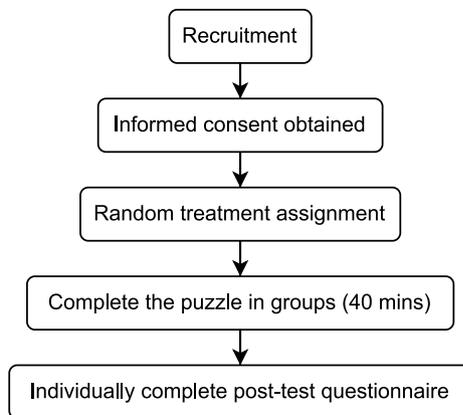

**Fig. 3.** After recruitment, independent participants were scheduled to attend a session. At the session, teams of three or four were randomly assigned one of the four voice assistant types. Then they completed an icebreaker activity before beginning the main puzzle task. After the puzzle task, they individually completed questionnaires, were paid, and finished the experiment. The research team then analyzed performance data at the team level and analyzed questionnaire response data at the individual level.

### 3.3. The audio-only agent

The agent communicated to participants through audio alone and displayed a black screen via the online video conferencing software. Although it would not respond to participant communication, an experimenter read a script to introduce the agent as "the Puzzle Master is an AI assistant that could offer new perspectives and information". The agent gave six clues over 40 min framed as guiding participants toward different milestones in the puzzle. We standardized each clue and gave them in the same order at the same time for every team. The clues and the agent's voice are the basis for our experimental manipulation, further described in Section 3.4.

The agent does not operate cooperatively with team members because it is automated and does not have autonomy. Hence, the teams in this experiment are not fully human–agent teams (at least one human working cooperatively with at least one autonomous agent (McNeese et al., 2018)). We find that this limitation is necessary to parse out the effect of human resemblance combined with contribution quality. With autonomous agents, confounds such as the co-creation of shared mental models (Schelble, Flathmann, McNeese, Freeman, & Mallick, 2022) and a user's familiarity with AI systems (Zhang et al., 2021) can alter performance and individual perceptions. By using an algorithm framed as an agent, we have better control of our internal validity.

### 3.4. Experimental manipulation

In a 2 × 2 design, we manipulated the agent's clue content and voice. First, its clues were either **helpful** or **unhelpful**. The helpful agent gave clues with true information such as, "The octopus rule relates to *separation* between the gems" (which is correct). The unhelpful agent gave clues with false information such as, "The octopus rule has to do with the *sizes* of the gems" (which is incorrect). This was intended to motivate teams to start looking for patterns within the octopus category. Second, its voice was either **human** or **robotic**. A female voice actor recorded the audio for the human voice condition, and a text-to-speech program generated the audio for the robotic voice condition. The robotic voice had a female timbre but was computer-generated with non-human prosody and paralanguage. The audio files are available online at https://github.com/samwestby/Voice-and-Contribution-in-HATs.

### 3.5. Treatment validation

We ran a pre-test validation to check if the robotic voice alone was perceived as non-human. We recruited 400 participants from Amazon Mechanical Turk. Eight variations of the treatment validation were run with eight different words for the robotic voice: *a machine, a computer, a robot, a digital assistant, an automated assistant, automation,* and *artificial intelligence*. 50 people each were assigned to a word. Participants listened to audio clips of our robotic agent's clues and then answered one question, "Is this voice a human or a machine?" on a seven-point Likert scale of agreement from "Definitely a machine" to "Definitely human". Note that "a machine" was only given to participants assigned to the "a machine" condition. We compared the proportion of responses for "Definitely [treatment word]" and "Very likely [treatment word]" to an expected proportion of 2/7 using a one-tailed Z-test. Participants were able to accurately label the voice as non-human, even when we manipulated the word referring to the robotic voice. The results for each word are as follows: a machine ($z = 3.111, p = 0.002$), a computer ($z = 3.677, p < 0.001$), a robot ($z = 2.263, p = 0.024$), a digital assistant ($z = 3.394, p = 0.001$), a virtual assistant ($z = 3.394, p = 0.001$), an automated assistant ($z = 2.828, p = 0.005$), automation ($z = 3.677, p < 0.001$), and artificial intelligence ($z = 2.546, p = 0.011$). We found no substantial difference between the specific labels (robot, computer, etc.). This validation confirms that the robotic voice was perceived as non-human.

### 3.6. Post-test questionnaire

After the puzzle, either after 40 min or when teams solved the puzzle ($n = 3$), participants completed a questionnaire. The questionnaire consisted of five sections, two of which are used in this paper and described below.

#### 3.6.1. The godspeed questionnaire

We used a subset of the Godspeed Questionnaire (Bartneck et al., 2009) to measure perceived anthropomorphism, perceived animacy, and perceived intelligence. Anthropomorphism is the attribution of human characteristics to non-human entities (Epley et al., 2007), animacy is the perception of acting on one's own accord (Bartneck et al., 2009), and perceived intelligence is the evaluation of the agent along dimensions of competence, knowledge, responsibility, intelligence, and sensibility (Bartneck et al., 2009). Participants answered five questions per category using a five-point semantic differential scale between opposing dyads. For example, "The Puzzle Master was *Artificial* versus *Lifelike*".

#### 3.6.2. Trustworthiness of the puzzle master

To measure trust in the Puzzle Master and perceptions of the Puzzle Master's contributions, we adapted a six-item questionnaire designed to measure trust in online avatars (Pan & Steed, 2016). Participants rated the following items on a five-point Likert scale from Strongly Disagree to Strongly Agree:

1. I trusted the Puzzle Master's clues.
2. I was well informed by the Puzzle Master.
3. The Puzzle Master gave helpful clues.
4. The Puzzle Master was competent.
5. The Puzzle Master was intentionally misleading the team.
6. The Puzzle Master was useful in helping the team solve the puzzle.

Notice that item (5) is reverse-coded. No participants failed this attention check.

### 3.7. Regressions

We conducted regressions to determine the effects of our manipulations and find interactions between contribution quality and the voice of the agent.

#### 3.7.1. Questionnaire

The questionnaire connects our manipulations of vocal human resemblance and contribution quality to perceptual outcomes. A multiple regression analysis with standard errors clustered by team was conducted using ordinary least squares (OLS) to fit a model for each questionnaire measure using the following equation,

$$\begin{aligned} \text{Measure} = &\beta_1 isHumanVoice \\ &+ \beta_2 isHelpfulClues \\ &+ \beta_3 isHumanVoice \times isHelpfulClues \\ &+ \epsilon \end{aligned} \quad (1)$$

where *isHumanVoice* and *isHelpfulClues* are binary {0, 1} variables depending on treatment assignment. *isHumanVoice* = 0 indicates an agent with a robotic voice instead of with a human voice. *isHelpfulClues* = 0 indicates an agent who gave unhelpful clues instead of helpful clues. "Measure" represents each section from the Post-Test Questionnaire (perceived anthropomorphism, perceived animacy, perceived intelligence, and perceived trustworthiness). Cluster-robust standard errors are used to account for clustering in standard errors between independent teams.

#### 3.7.2. Team performance

Team performance was measured as the number of milestones a team achieved (min. 1, max. 7). We calculated a multiple linear regression using OLS to predict team performance based on the treatment category with the following equation,

$$\begin{aligned} \text{Team Performance} = &\beta_1 isHumanVoice \\ &+ \beta_2 isHelpfulClues \\ &+ \epsilon \end{aligned} \quad (2)$$

We tested adding an interaction term between *isHumanVoice* and *isHelpfulClues*, but removed it due to insignificance ($p = 0.713$). Performances at two different time points were used. The first was the number of milestones achieved at the three-quarter time point (thirty min), and the second was the number of milestones achieved at the final time point (forty min).

## 4. Results

Of the sixty-nine participants, all sixty-nine completed the experiment and questionnaire. For each questionnaire construct, we measured internal consistency using Cronbach's Alpha and found that all constructs were reliable (*anthropomorphism* $\alpha = 0.79$, *animacy* $\alpha = 0.81$, *perceived intelligence* $\alpha = 0.84$, and *perceived trustworthiness* $\alpha = 0.93$).

First, we will address how an agent's voice combines with an agent's contribution to affect human perceptions (see Figs. 5 and 6).

### 4.1. Perceived anthropomorphism and perceived animacy

The human or robotic voice manipulation altered anthropomorphism ($p = 0.033$) and animacy ($p = 0.042$). Counterintuitively, the human voice was not always perceived as more human than the robotic voice. Agents with a human voice that gave helpful clues were seen as less anthropomorphic and animated than agents with a human voice and unhelpful clues as shown in Fig. 4a and Fig. 4b. **H1.1**, that agents with human voices will be perceived as more anthropomorphic and animated than agents with robotic voices, is not supported.

Agent voice, agent clue quality, and the interaction between the two were all significant factors for regressions on anthropomorphism and animacy. The interaction term ("Human Voice × Helpful Clues" in Table 2) for perceived anthropomorphism and animacy is negative, meaning voice type flips the effect of an agent's helpfulness. Helpfulness increases the perceived anthropomorphism and animacy of robotic-voiced agents whereas helpfulness decreases the perceived anthropomorphism and animacy of human-voiced agents. The relationship between an agent's human resemblance and perceived anthropomorphism and animacy is not always positive. We accept hypothesis **H1.2**, that voice type will moderate the effect of contribution quality on perceived anthropomorphism and perceived animacy.

### 4.2. Perceived intelligence and perceived trustworthiness

The voice of the agent had no significant effect on the perceived intelligence or perceived trustworthiness of the agent (*intell.*: $p = 0.183$; *trust*: $p = 0.405$), although the full regression predicted perceived intelligence (Adj. $R^2 = 0.219$, $F(3, 65) = 6.51$, $p = 0.003$) and perceived trustworthiness (Adj. $R^2 = 0.733$, $F(3, 65) = 50.4$, $p < 0.001$). There was no evidence to support that voice type affects perceived intelligence or perceived trustworthiness. We did find that agents giving unhelpful clues were rated as significantly less trustworthy ($p < 0.001$, Table 2, model 3) and less intelligent ($p < 0.001$, Table 2, model 4) than agents giving helpful clues as shown in Fig. 5 and Fig. 4c respectively. In shorter studies with less interaction between humans and agents, researchers may continue to find evidence for the effect of an agent's attributes. In this study, we find that perceptions of intelligence and trustworthiness are dependent on contribution and not human resemblance. We accept hypotheses **H2.1** and **H3.1**. The interaction of voice and contribution quality did not impact trustworthiness, so hypothesis **H3.2** is not supported. Participants did not trust unhelpful human agents any different than unhelpful robotic agents.

**Table 2**
Results for multiple regressions of questionnaire items versus treatment with cluster-robust standard errors to account for clustering in standard errors between independent teams following Eq. (1). This table also includes results for multiple regressions of team performance versus treatment following Eq. (2).

| Dependent variable: | Questionnaire item (Individual level) | | | | Performance (Team level) | |
|---|---|---|---|---|---|---|
| | Anthro. (1) | Animacy (2) | Intell. (3) | Trust (4) | 30 min. (5) | 40 min. (6) |
| Human voice | 2.53* | 3.47* | 2.41 | 1.94 | −0.20 | −0.20 |
| | (1.19) | (1.71) | (1.81) | (2.33) | (0.52) | (0.70) |
| Helpful clues | 3.53** | 3.95** | 6.52*** | 15.64*** | 1.40* | 1.20 |
| | (1.19) | (1.14) | (1.68) | (1.53) | (0.52) | (0.70) |
| Human voice × Helpful clues | −4.88** | −6.48** | −4.05 | −3.28 | | |
| | (1.67) | (2.12) | (2.29) | (2.78) | | |
| Adj. $R^2$ | 0.06 | 0.08 | 0.22 | 0.73 | 0.22 | 0.05 |
| Num. obs. | 69 | 69 | 69 | 69 | 20 | 20 |

*** $p < 0.001$.
** $p < 0.01$.
* $p < 0.05$.

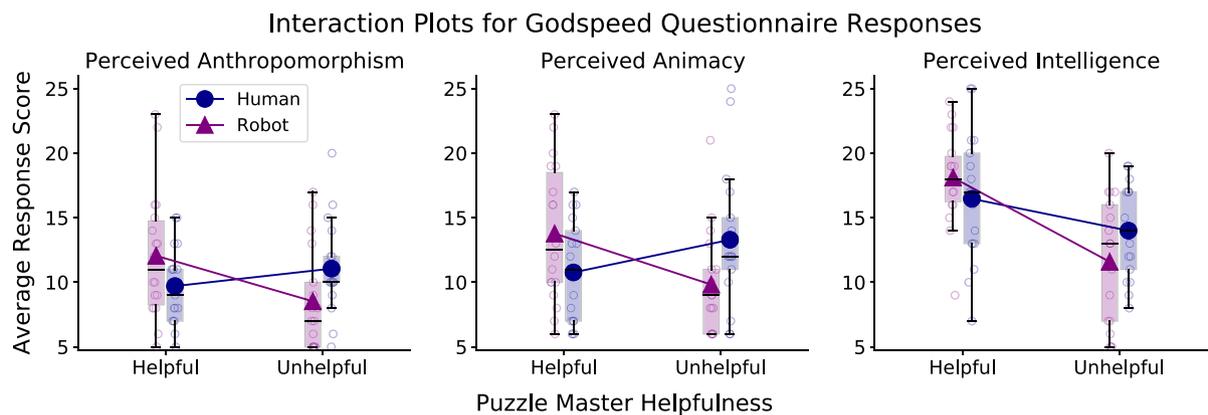

**Fig. 4.** These three plots show the presence or absence of interactive effects caused by the voice or contribution of the agent for individual questionnaire responses. For both perceived anthropomorphism and animacy, the agent's voice flips the effect of helpfulness. In some cases, being helpful makes the agent seem more human, and in other cases it makes the agent seem less human. Fig. 4C shows that perceptions of agent intelligence were only significantly changed by the contribution quality of the agent teammate.

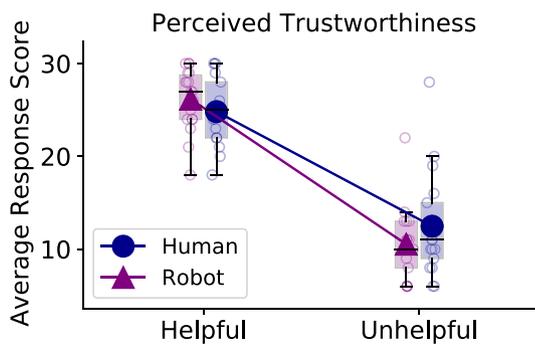

**Fig. 5.** This plot shows that agent voice does not affect perceived trustworthiness. Agent contribution was the only factor that caused differences in perceived trustworthiness ($p < 0.001$, Table 2 Model 4).

### 4.3. Team performance

Here we address the effect of voice and contribution on performance. Treatment assignment caused differences in team performance at the three-quarter time point ($F(2, 18) = 3.70, p = 0.047$, Table 2, model (5)). With an adjusted $R^2$ of 0.221, treatment assignment predicted 22.1% of the variance in a team's performance at the three-quarter mark. Of the two treatment variables, only agent helpfulness was a significant factor (voice, $p = 0.705$; **helpfulness, $p = 0.015$**) indicating agent helpfulness predicts team performance at the three-quarter mark while agent voice type does not. This difference was not significant by the end of the puzzle ($F(2, 18) = 1.52, p = 0.247$, Adj. $R^2 = 0.05$, Table 2, model 6). At the three-quarter mark, teams were less likely to know whether the agent was helpful or unhelpful than at the end of the puzzle. Upon realizing the agent was unhelpful, several teams in the unhelpful agent condition showed fast achievement of milestones before the 40-min stopping time. Teams in the helpful agent condition correctly solved 1.40 ± 0.520 more milestones than teams in the unhelpful condition irrespective of the agent's voice. These differences are shown in Fig. 6. These results partially support **H4.1** that teams with a helpful agent will perform better than teams with an unhelpful agent. Because no significant differences were found between the performances of teams with human-voiced agents and robotic-voiced agents, we conclude that voice type does not affect a team's performance. Hypothesis **H4.2** is not supported.

### 5. Discussion

The results presented here improve our understanding of how human resemblance and contribution quality affect perceptions of voice assistants and performance in human–agent teams. We manipulated the voice and helpfulness of a voice assistant in 20 human–agent teams. First, the human resemblance of an agent's voice had a negative interaction with an agent's helpfulness, resulting in flipped perceived anthropomorphism and animacy. A helpful robotic-voiced agent was perceived as more human than an unhelpful robotic-voiced agent, and conversely, a helpful human-voiced agent was perceived as less human than an unhelpful human-voiced agent. Second, varying the agent's voice did not change perceived intelligence, trust in the agent, or team performance. Third, the helpful vs. unhelpful condition significantly affected perceived intelligence, trust in the agent, and team performance. In the following, we will discuss how these results are important for designers and researchers.

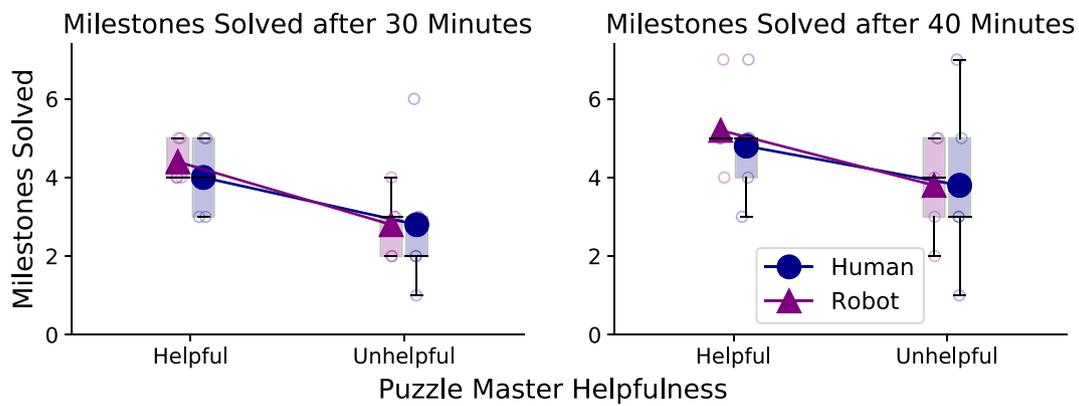

**Fig. 6.** These plots show the effect of treatment categories on team performance. After 30 min which was 75% of the allotted time, treatment significantly predicted 22.1% of the variance in team performance ($p = 0.047$, Table 2 Model 5). Only agent contribution had a significant coefficient in the regression equation ($p = 0.015$, Table 2 Model 5) which means agent voice did not affect performance. After 40 min at the end of the task, there was no significant effect of treatment on performance ($p = 0.247$, Table 2 Model 6).

*5.1. Perceived anthropomorphism and animacy*

The factors of perceived anthropomorphism have been the subject of conflicting research. Faulty and less helpful teammates, when compared to flawless and more helpful teammates, have been rated higher (Salem et al., 2013), lower (Salem et al., 2015), and the same (Mirnig et al., 2017) in perceived anthropomorphism. This conflict led us to **H1.2**, that voice type moderates the effect of contribution quality on perceived anthropomorphism.

The observed negative interaction between human resemblance in voice and an agent's contribution quality shows that increasing vocal human resemblance does not guarantee higher anthropomorphization, hence **H1.1** is not supported. The interaction may be due to participants' expectations of the agent, which could be either met or violated. Humans may approach social situations with a set of scripts specifically developed for the subject of their interaction as stated by CASA (Gambino et al., 2020; Nass & Moon, 2000). Here, human-voiced agents may activate different scripts than robotic-voiced agents, producing different perceptions and expectations. The robotic-voiced agent is seen as more human when it gives helpful clues, but the human-voiced agent is seen as more human when it gives unhelpful clues. Participants may expect a human-voiced agent to act more like them. The puzzle was difficult for most teams, and only three of the twenty teams completed it before time expired. It is plausible that participants viewed the puzzle as beyond the human-voiced agent's abilities. The agent should struggle just like the humans would struggle. On the other hand, they may have expected the robotic-voiced agent to be more of an oracle with insights beyond human abilities. This also explains when the agents were seen as less human. An all-knowing human-voiced agent did not fit with the participant's expectations for what a human should know. An unhelpful robotic-voiced agent also did not fit with established social scripts for interactions.

Perceived animacy shows the same negative interaction as perceived anthropomorphism. This tells us that voice interacted with contribution quality to alter a person's belief that the voice assistant acted on its own will. When the agent contradicts a user's behavioral expectations, it may be seen as "soulless". Participants see the agent as random, algorithmic, and not autonomous. The agent becomes incapable of following the expected social scripts.

*Implications.* Designers and researchers should be aware of the fickle nature of perceived anthropomorphism and animacy. Contrary to Moussawi and Benbunan-Fich (2021) and Ferstl et al. (2021), altering the human resemblance of an agent's voice alone is not enough to control perceived anthropomorphism and perceived animacy. An agent's behavior can change what seems human (Jung et al., 2015; Mirnig et al., 2017). Our results show that measures of anthropomorphism and animacy are more complicated than a linear scale of human resemblance. An agent's contribution quality can cause agents with human voices to be rated lower in perceived anthropomorphism and animacy than agents with robotic voices. Those attempting to control or optimize perceived anthropomorphism and animacy may not be able to draw conclusions without testing their agent in the wild. The human resemblance of an agent's voice may outweigh or underweigh the impact of an agent's actions.

*5.2. Intelligence and trust*

The voice of the agent did not affect perceived intelligence or perceived trustworthiness, which aligns with (Moussawi & Benbunan-Fich, 2021). Perceived intelligence and perceived trustworthiness were entirely predicted by the helpful/unhelpful treatment condition. Low-quality information may have caused this result. We noticed two ways that teams identified when an agent was unhelpful. If a team solved a milestone before an agent gave an unhelpful clue for the same milestone, then they became skeptical of the agent. Other teams became skeptical of the unhelpful agent after repeatedly failing to apply its unhelpful clues. They stopped trusting the agent's information.

Although we did not measure perceived trust and intelligence at the beginning of the task, it is clear that team members calibrated their ratings to the actual merit of the agent by the end of the task. Only agent helpfulness, not agent voice, caused changes in perceived intelligence and perceived trustworthiness (Table 2). We accept hypotheses **H2.1** and **H3.1**. This shows that perceived intelligence and trustworthiness can be objective once participants have had sufficient interaction with the agent. Humans successfully learned when agents were unhelpful and rated the agent as such. This aligns with prior work on trust calibration (Demir et al., 2021; Lee & See, 2004) and on the interaction between helpfulness and human resemblance on trust (Kulms & Kopp, 2019). These are important factors for long-term usage (Parasuraman & Riley, 1997).

*Implications.* Designers and researchers who want to increase the perceived intelligence and trustworthiness of agents should improve agent functionality before fine-tuning their voices. We find that participants saw beyond the differences in voice and objectively rated the agent's intelligence and trustworthiness based on ability and contribution. As a user interacts with the agent, they learn its value. They learn whether or not the tool will help them complete their task. This is the long-term reason to use voice assistants. People prefer skilled agents (Zhang et al., 2021). Although prior work finds that physical characteristics may impact initial adoption, voice assistants who cannot contribute will be seen as less intelligent, not trustworthy, and thus not used.

*5.3. Performance*

Teams with a helpful agent solved significantly more milestones than teams with an unhelpful agent after 30 min. Looking deeper, that influence was not from whether the agent had a human or robotic voice. It was only from whether the agent gave helpful clues or unhelpful clues. Despite (Visser et al., 2016) finding humans trusting unhelpful human-like agents longer than unhelpful non-human-like agents, we do not. Hypothesis **H3.2** that participants will trust unhelpful human agents more than unhelpful robotic agents, is not supported. Also **H4.2**, teams with an unhelpful human-voiced agent will perform worse than teams with an unhelpful robotic-voiced agent, is not supported. There may be a possible subtle effect, but our analyses do not have enough power to detect it.

*Implications.* Although treatment assignment did not significantly influence final team performance, this result contextualizes design priorities for performance-focused human–agent teams. In tasks similar to this experiment, designers and researchers should fully optimize agent contribution before making marginal gains with voice type and physical characteristics like voice. A helpful agent will create a higher performing HAT than a less helpful agent, regardless of form. With time, humans can calibrate their interpretations of the actions that an agent takes (Demir et al., 2021). They will move past short-term subjective effects caused by the agent's form. Function matters more than form.

## 6. Conclusion

In this study, we investigated the impact of an agent's voice and contribution quality on human perceptions and team performance in human–agent teams (HATs). This sheds light on the higher-order interaction of multiple factors in agent design. We show that a voice assistant's voice and contribution combine in counter-intuitive ways to alter perceived anthropomorphism and animacy. This interaction does not apply to other measures. Team performance, perceived agent intelligence, and agent trustworthiness were all dependent on a voice assistant's contribution, not voice.

*Limitations.* Two major factors that we did not manipulate in our study are task type and environment. All teams were on a video-conferencing call, using identical laptops in unfamiliar rooms, and working with a team of strangers to solve a difficult puzzle. These factors could have influenced perceptions and performance in ways that we did not account for. For example, different task types or environments might lead to different levels of perceived anthropomorphism, animacy, or team performance. Additionally, we did not vary the level of autonomy of the agent in our study. Our agent provided clues based on a predetermined script, without the ability to adapt to the team's progress or provide personalized assistance. A more autonomous agent, capable of learning and adapting to the team's needs, might have different effects on perceptions and performance.

We did not measure the usability and likability of the agent. Usability is a major problem to solve, and it is closely tied to likability. Although these two measures were out of the scope of this project, future studies that include them can provide a more comprehensive understanding of human–agent interaction. Finally, our regressions on team performance had five observations per treatment. This is under powered for our effect sizes. It is possible that we did not have enough observations to uncover an effect in a reasonable confidence interval.

We acknowledge that the sample may not fully represent the broader population of voice assistant users, especially those from other geographic regions or with different technological backgrounds. The reliance on convenience sampling methods and recruitment in the Northeast United States could introduce sampling bias, potentially limiting the generalizability of our findings. To mitigate this, we recruited participants from a variety of community settings to increase demographic diversity. Future studies should aim to include participants from more diverse geographic locations and a wider range of technological familiarity.

Despite these limitations, our study contributes to the understanding of human–agent teams by highlighting the delicate nature of perceived anthropomorphism and perceived animacy as well as the objective nature of performance, perceived intelligence, and trust.

**CRediT authorship contribution statement**

**Samuel Westby:** Writing – review & editing, Writing – original draft, Visualization, Investigation, Formal analysis. **Richard J. Radke:** Writing – review & editing, Methodology, Conceptualization. **Christoph Riedl:** Writing – review & editing, Methodology, Conceptualization. **Brooke Foucault Welles:** Writing – review & editing, Methodology, Conceptualization.

**Declaration of Generative AI and AI-assisted technologies in the writing process**

During the preparation of this work, the authors used ChatGPT to improve the language and readability of this paper. After using this tool, the authors reviewed and edited the content as needed and take full responsibility for the content of the publication.

**Declaration of competing interest**

The authors declare that they have no known competing financial interests or personal relationships that could have appeared to influence the work reported in this paper.

**Acknowledgments**

We gratefully acknowledge the contributions of Kristen Flaherty, Robin Lange, and Eri Lee. This work was supported by the Army Research Laboratory, United States [Grant W911NF-19-2-0135].